\documentclass[12pt]{iopart}
\usepackage{iopams}  
\usepackage{amssymb}
\usepackage{setstack}  
\usepackage{graphicx}
\usepackage{bm}
\usepackage{amsbsy}
\usepackage{amstext}
\usepackage{iopams}
\usepackage[T1]{fontenc}
\usepackage[latin9]{inputenc}

\begin{document}

\title{Three approaches for representing Lindblad dynamics by a matrix-vector notation}

\author{Morag Am-Shallem, Amikam Levy, Ido Schaefer and Ronnie Kosloff}
\address{
Fritz Haber Research Center and the Institute of Chemistry,
The Hebrew University, Jerusalem 91904, Israel}

\begin{abstract}
Markovian dynamics of open quantum systems are described
by the L-GKS equation, known also as the Lindblad equation. 
The equation is expressed by means of left and right matrix multiplications. 
This formulation hampers numerical implementations. 
Representing the dynamics by a matrix-vector notation overcomes this problem.
We review three approaches to obtain such a representation. 
The methods are demonstrated  for a driven two-level system subject to spontaneous emission. 
\end{abstract}

\section{Introduction}

An open system is a system that interacts with its environment. A
full description has to account for all the degrees of freedom (DOF) of
the entire system and its environment. Usually, only the system DOF
are of interest. A reduced description attempts to describe only the
system DOF explicitly, while the environment DOF are integrated
out and affect the description implicitly. The goal is to reduce the
description to a small number of variables and obtain a practical way
to treat the system \cite{kraus1971general,davies1976quantum,gardiner1985handbook}.

Open systems are often described by a stochastic process which in
many cases becomes a simple Markov process. 
In brief, a Markov
process is a stochastic process with a short time memory, i.e., the
process state depends solely on the present state. Mathematically
it can be constructed as the Chapman-Kolmagorov equation for the conditional
joint probability \cite{gardiner1985handbook}. For a Markov process the
probability distribution $p_{t}(x)$ on a certain space, either real-space
or phase-space, which could be continuous or discrete, follows the
differential equation 
\begin{equation}
\frac{d}{dt}p_{t}(x)=\mathcal{L}p_{t}(x),\label{eq:master_class}
\end{equation}
The formal solution for Eq.(\ref{eq:master_class}) is given by,
\begin{equation}
p_{t}(x)=\Lambda_{t}p(x),\qquad\Lambda_{t}=e^{\mathcal{L}t},\qquad t\geq0,\label{eq:L_formal_solution}
\end{equation}
where, without loss of generality, we define the initial time to be zero. 
The one-parameter family of maps$\{\Lambda_{t},t\geq0\}$ is
a \textit{semigroup }with the generator $\mathcal{L}$. 
The term \textit{semigroup} implies that this family of maps does not form a full group.
It lacks the negative range of the parameter $t$, which implies that
the inverse property required by a group is missing. 
Physically, this property is the manifestation of irreversible dynamics which
allows us to distinguish the future from the past. 
The map $\Lambda_{t}$ is a positive map that satisfies the composition rule (Markov property)
$\Lambda_{t+s}=\Lambda_{t}\Lambda_{s}\quad t,s\geq0$, and preserves
normalization of the probability density.

In the quantum scenario several modifications have to be made. 
The probability distribution $p_{t}(x)$ is replaced by the density matrix
$\boldsymbol{\mathrm{\hat{\rho}}}(t)$.
The property of positivity has to be strengthened to complete positivity.
The dynamics follows the quantum master equation: 
\begin{equation}
\label{eq:master_quant}
\frac{d}{dt}\boldsymbol{\mathrm{\hat{\rho}}}(t)=\mathcal{L}\boldsymbol{\mathrm{\hat{\rho}}}(t).
\end{equation}
This is a direct consequence of the presence of entangled states
\cite{alicki2001quantum}. 
To summarize, the quantum dynamical \textit{semigroup} is a continuous one-parameter family of maps $\{\Lambda_{t},t\geq0\}$,
that satisfies \cite{alicki2007qds}:
\begin{enumerate}
\item $\Lambda_{t}$ is complete positive; 
\item $\Lambda_{t}$ is trace preserving;
\item $\Lambda_{t+s}=\Lambda_{t}\Lambda_{s}\quad t,s\geq0$ \textit{semigroup
}(Markov) property;
\item $\Lambda_{t}$ is strongly continuous. 
\end{enumerate}
\par
Lindblad as well as Gorini, Kossakowski and Sudarshan (L-GKS)
introduced the most general form of the quantum dynamical \textit{semigroup} 
generator $\mathcal{L}$ that satisfies these requirements 
\cite{lindblad1976generators,gorini1976completely}.
In the Lindblad form the Markovian master equation reads:
\begin{equation}
\begin{array}{rcccc}
\frac{d}{dt}\boldsymbol{\mathrm{\hat{\rho}}}(t)=\mathcal{L}\boldsymbol{\mathrm{\hat{\rho}}} & = & -\frac{i}{\hbar}\left[\boldsymbol{\mathrm{\hat{H}}},\boldsymbol{\mathrm{\hat{\rho}}}\right] & + & \sum_{i}\gamma_{i}\left(\boldsymbol{\mathrm{\hat{A}}}_{i}\boldsymbol{\mathrm{\hat{\rho}}}\boldsymbol{\mathrm{\hat{A}}}_{i}^{\dagger}-\frac{1}{2}\left\lbrace \boldsymbol{\mathrm{\hat{A}}}_{i}^{\dagger}\boldsymbol{\mathrm{\hat{A}}}_{i},\boldsymbol{\mathrm{\hat{\rho}}}\right\rbrace \right)\\
 & \equiv & \mathcal{L}_{H}(\boldsymbol{\mathrm{\hat{\rho}}}) & + & \mathcal{L}_{D}(\boldsymbol{\mathrm{\hat{\rho}}}).
\end{array}\label{eq:LGKS_schrod}
\end{equation}
 Here, $\boldsymbol{\mathrm{\hat{H}}}$ is the effective Hamiltonian
of the system, $\gamma_{i}$ are positive rates, and $\{\boldsymbol{\mathrm{\hat{A}}}_{i}\}$
are operators belonging to the Hilbert space of the system. We use the
notation $\mathcal{L}_{H}$ to represent the unitary part of the dynamics,
and $\mathcal{L}_{D}$ to represent the dissipative part. $\mathcal{L}$,
$\mathcal{L}_{H}$ and $\mathcal{L}_{D}$ are linear operators that
operate on the density matrix, usually referred to as \emph{super-operators}. 

The operation of the super-operator $\Lambda_{t}=e^{\mathcal{L}t}$
on the density matrix could be understood as repetitive operations of
the super-operator $\mathcal{L}$ as in the Taylor expansion:

\begin{equation}
e^{\mathcal{L}t}\boldsymbol{\mathrm{\hat{\rho}}}\equiv\sum_{k}\frac{1}{k!}\mathcal{L}^{k}\boldsymbol{\mathrm{\hat{\rho}}}t^{k}=\boldsymbol{\mathrm{\hat{\rho}}}+\mathcal{L}\boldsymbol{\mathrm{\hat{\rho}}}t+\frac{1}{2}\mathcal{L}^{2}\boldsymbol{\mathrm{\hat{\rho}}}t^{2}+\ldots\label{eq:L_exponential}
\end{equation}
Typically, the resulting dynamics of the system observables (expectation
values and other correlation functions) $c(t)$ will have the analytical
form of sum of decaying oscillations\footnote{
	There are special cases where the super-operator is not diagonalizable.
	In such cases, known as exceptional points, the exponential $e^{\lambda t}$
	is multiplied by a polynomial of $t$. A study of exceptional points
	in L-GKS system can be found in Refs \cite{morag2014EPbloch,morag2015EPatomic}.
}:
\begin{equation}
c(t)=\sum_{m}d_{m}e^{\lambda_{m}t},\label{eq:sum_exp}
\end{equation}
Here, $\lambda_{m}$ are the exponential coefficients and $d_{m}$
are the associated amplitudes, both can be complex. We may divide $\lambda_{m}$
into its real and imaginary parts, $\lambda_{m}=-\alpha_{m}+i\omega_{m}$,
with $\alpha_{m}\ge0 \in \mathbb{R}$ as the decay rates and $\omega_{m}\in \mathbb{R}$ as the
oscillation frequencies. The coefficients $\lambda_{m}$ are the eigenvalues
of the super-operator $\mathcal{L}$, obtained by the eigenvalue equation:
\begin{equation}
\mathcal{L}\boldsymbol{\mathrm{\hat{\sigma}}}_{m}=\lambda_{m}\boldsymbol{\mathrm{\hat{\sigma}}}_{m}.\label{eq:eig_of_L}
\end{equation}
These eigenvalues can be used for the analysis of the L-GKS dynamics.

As noted above, the dynamics can be investigated by exponentiation of
the super-operator $\mathcal{L}$, Eq. (\ref{eq:L_exponential}),
or by its eigenvalues, Eq. (\ref{eq:eig_of_L}). The exponentiation
and the eigenvalue problem of the (linear) super-operator $\mathcal{L}$
are well defined. However, they are not suitable for numerical calculations.
Calculations of the exponentiation and the eigenvalue equation of linear operators
can be done by common numerical techniques if the linear operator
is represented by a matrix. 
Therefore, a preferred representation of the dynamics, Eq. (\ref{eq:LGKS_schrod}), 
is in a matrix-vector notation. 
This means that we are looking for a matrix $L$ and a vector
$\vec{r}_{s}$ such that the dynamics are expressed as 
\begin{equation}
\frac{d}{dt}\vec{r}_{s}=L\vec{r}_{s}.\label{eq:mat_vec_not}
\end{equation}
In this representation, the vector $\vec{r}_{s}$ represents the state of the system, 
or some information about it, e.g. a set of expectation values.
Next, we describe three approaches for such a representation,
and demonstrate them for a case of a driven two-level system with relaxation.

\section{Matrix-vector representations}

Suppose the density matrix $\boldsymbol{\mathrm{\hat{\rho}}}$ is
an $n\times n$ matrix (if $\boldsymbol{\mathrm{\hat{\rho}}}$ is
a function of continuous variables, e.g. $\boldsymbol{\mathrm{\hat{\rho}}}(r,r')$,
these variables have to be discretized). 
The set of all $n\times n$
matrices form a linear space of dimension $n^{2}$. 
Under  appropriate conditions, 
this linear space can have a Hilbert space construction,
using the scalar product defined as 
\[
\left(\boldsymbol{\mathrm{\hat{\rho}}}_{1},\boldsymbol{\mathrm{\hat{\rho}}}_{2}\right)=\text{Tr}\left\{ \boldsymbol{\mathrm{\hat{\rho}}}_{1}^{\dagger}\boldsymbol{\mathrm{\hat{\rho}}}_{2}\right\} .
\]
Such a Hilbert space is called a Liouville space 
(also known as the Hilbert-Schmidt space). 
With such a construction we consider $\boldsymbol{\mathrm{\hat{\rho}}}$
as an $n^{2}$ vector. 
Similarly, we consider the super-operator
${\cal L}$, which is an operator operating on elements in this
linear space, as an $n^{2}\times n^{2}$ matrix. 

The above observation is the first step towards the representation
we seek. 
In the following, we describe three approaches
that use this concept to introduce such representation:
\begin{enumerate}
\item \emph{Vec-ing the density matrix} is the most natural way to construct
an $n^{2}$ vector for the density matrix, and a suitable $n^{2}\times n^{2}$
matrix for the super-operator.
\item \emph{The Arnoldi method} approximates a large matrix in smaller dimensions,
enabling simpler numerical calculations.
\item With \emph{the Heisenberg picture} of the L-GKS equation we can search
for a representation with a dimension smaller than $n^{2}$. 
\end{enumerate}
In the following, we describe these three approaches. Each of these
approaches will be demonstrated in the case of the two-level
system.

\subsection{Vec-ing the density matrix\label{sec:vec_theory}}

In this method, known as vec-ing \cite[Chapter 4]{machnes14surprising,roger1994topics},
the $n\times n$ density matrix $\boldsymbol{\mathrm{\hat{\rho}}}$
is flattened into an $n^{2}$ vector $\vec{r}$ . This flattening
is done by ordering the columns of $\boldsymbol{\mathrm{\hat{\rho}}}$
one below the other, so the $(a,b)$ entry of the matrix $\boldsymbol{\mathrm{\hat{\rho}}}$
is the ${(b-1)n+a}$ entry of the vector $\vec{r}$. This is equivalent
to choosing the representation basis as the set of matrices with all-zero
entries, except one.

The next task is to find the suitable matrix that will represent the
operation of the super-operator $\mathcal{L}$ on the density matrix.
We make the following observations \cite{machnes14surprising,roger1994topics}:
\begin{enumerate}
\item A left multiplication of the matrix $\boldsymbol{\mathrm{\hat{\rho}}}$
by an $n\times n$ matrix $A$, i.e. $A\boldsymbol{\mathrm{\hat{\rho}}}$,
is equivalent to an operation on the vector $\vec{r}$ by the $n^{2}\times n^{2}$
matrix $I\otimes A$, where $I$ is the $n\times n$ identity matrix,
and $\otimes$ is the Kronecker direct product.
\item Similarly, a right multiplication of the matrix $\boldsymbol{\mathrm{\hat{\rho}}}$
by an $n\times n$ matrix $B$, i.e. $\boldsymbol{\mathrm{\hat{\rho}}}B$,
is equivalent to an operation on the vector $\vec{r}$ by the $n^{2}\times n^{2}$
matrix $B^{T}\otimes I$. Here $T$ denotes the transpose of the matrix.
\item Finally, a combination of left and right matrices multiplication, $A\boldsymbol{\mathrm{\hat{\rho}}}B$,
is equivalent to an operation on the vector $\vec{r}$ by the $n^{2}\times n^{2}$
matrix $B^{T}\otimes A$.
\end{enumerate}
The L-GKS super-operator is a sum of such right and left multiplications.
Therefore, the construction of the $n^{2}\times n^{2}$ matrix representation
for the L-GKS generator has the parts as follows;  
for the commutator:
\[
\left[\boldsymbol{\mathrm{\hat{H}}},\boldsymbol{\mathrm{\hat{\rho}}}\right]\rightarrow\left(I\otimes\boldsymbol{\mathrm{\hat{H}}}-\boldsymbol{\mathrm{\hat{H}}}^{T}\otimes I\right)\vec{r}.
\]
For the dissipative part: 
\[
\begin{array}{rcl}
\boldsymbol{\mathrm{\hat{A}}}_{i}\boldsymbol{\mathrm{\hat{\rho}}}\boldsymbol{\mathrm{\hat{A}}}_{i}^{\dagger} & \rightarrow & \left(\left(\boldsymbol{\mathrm{\hat{A}}}_{i}^{\dagger}\right)^{T}\otimes\boldsymbol{\mathrm{\hat{A}}}_{i}\right)\vec{r}\\
\boldsymbol{\mathrm{\hat{A}}}_{i}^{\dagger}\boldsymbol{\mathrm{\hat{A}}}_{i}\boldsymbol{\mathrm{\hat{\rho}}} & \rightarrow & \left(I\otimes\boldsymbol{\mathrm{\hat{A}}}_{i}^{\dagger}\boldsymbol{\mathrm{\hat{A}}}_{i}\right)\vec{r}\\
\boldsymbol{\mathrm{\hat{\rho}}}\boldsymbol{\mathrm{\hat{A}}}_{i}^{\dagger}\boldsymbol{\mathrm{\hat{A}}}_{i} & \rightarrow & \left(\left(\boldsymbol{\mathrm{\hat{A}}}_{i}^{\dagger}\boldsymbol{\mathrm{\hat{A}}}_{i}\right)^{T}\otimes I\right)\vec{r}.
\end{array}
\]
Then we write 
\begin{equation}
L=I\otimes\boldsymbol{\mathrm{\hat{H}}}-\boldsymbol{\mathrm{\hat{H}}}^{T}\otimes I+\sum_{i}\gamma_{i}\left(\left(\boldsymbol{\mathrm{\hat{A}}}_{i}^{\dagger}\right)^{T}\otimes\boldsymbol{\mathrm{\hat{A}}}_{i}-\frac{1}{2}\left(I\otimes\boldsymbol{\mathrm{\hat{A}}}_{i}^{\dagger}\boldsymbol{\mathrm{\hat{A}}}_{i}+\left(\boldsymbol{\mathrm{\hat{A}}}_{i}^{\dagger}\boldsymbol{\mathrm{\hat{A}}}_{i}\right)^{T}\otimes I\right)\right),\label{eq:L_vec_ing}
\end{equation}
and represent Eq. (\ref{eq:LGKS_schrod}) as 
\[
\frac{d}{dt}\vec{r}=L\vec{r}
\]
as desired. 

The mapping of the density matrix into a density vector yields in a dramatic increment in the dimension of the problem, which becomes $n^2$ instead of $n$. 
This yields unfavorable scaling of the desired computations with $n$: 
\begin{itemize}
\item \emph{Eigenvalue approach.}
Computation of the complete eigenvalue spectrum of $L$ is performed via the diagonalization of $L$. 
Diagonalization of a matrix scales as the cube of its dimension. 
Hence, the diagonalization of $L$ scales as $n^{6}$. 
\item \emph{Exponentiation methods.}
The exponentiation of the matrix for time propagation, $e^{Lt}$, 
can be  computed via various ways \cite{moler1978nineteen_25}.
Remarkably, two branches are of interest:
\begin{enumerate}
\item Directly employing the diagonalization of $L$.
\item By numerical approximations, which usually involve matrix-matrix multiplications. 
\end{enumerate}
Both diagonalization and matrix-matrix multiplications scale as the cube of the matrix dimension.
Therefore, the overall scaling of the exponentiation is also $n^{6}$.
 
The calculation cost of the operation of the exponential $e^{Lt}$ on an initial vector $\vec{r}_0$, i.e. $e^{Lt}\vec{r}_{0}$, 
can be reduced by employing matrix-vector multiplications, and therefore scales as $n^{4}$ \cite{AlMohy2011expmv}. 
\end{itemize}
For systems larger than a few degrees of freedom, such computations are expensive, 
and become practically impossible for systems larger than a few hundreds DOF.

The scaling problem suggests that we have to look for approaches that use
a smaller number of dimensions. 
The following two approaches address this issue. 
The Arnoldi method uses a small-dimension approximation of a large matrix. 
The operator representation seeks for a small subset
of variables that are sufficient to describe the quantities of interest.
These two approaches are described in the next two sections.

\emph{Remark}: The density matrix $\boldsymbol{\mathrm{\hat{\rho}}}$ is
hermitian. Therefore there are only $n(n+1)/2$ unique entries and
not $n^{2}$. This fact can be used to reduce the size of the vectors
and matrices, known as a half-vectorization \cite[Chapter 11]{abadir2005matrix}.
However, we will not discuss this here.

\subsection{Arnoldi method\label{sec:arnoldi_theory}}

The Arnoldi method is a method to approximate a large matrix $A$ in a smaller dimension \cite{trefethen1997numerical}. 
This is done by choosing an appropriate
set of a small number of vectors, which should be representative of
the relevant subspace for a specific problem. Then the desired matrix
is represented in the reduced subspace which is spanned by the chosen
vectors. The method starts with an initial vector $\vec{v}$ and creates
set of $K+1$ vectors by the repetitive operation of the matrix $A$:
$\{\vec{v},\allowbreak
A\vec{v},\allowbreak
A^{2}\vec{v},\allowbreak 
\ldots,\allowbreak 
A^{K}\vec{v}\}$. 
Then an orthonormal set is generated from this set by the Gram-Schmidt process.
This orthonormal vectors set spans a subspace with dimension $K+1$,
and the matrix $A$ is represented in this subspace by a $(K+1)\times(K+1)$
matrix. 
This smaller matrix can be
used for the efficient evaluation of functions of the matrix $A$, e.g.
the exponential \cite{saad1992analysis_krylov} 
or the eigenvalues \cite{arnoldi1951principle}. 

In our case we try to approximate the linear super-operator $\mathcal{L}$
by a matrix which is smaller than $n^{2}\times n^{2}$. Conceptually,
we start with the initial density matrix $\boldsymbol{\mathrm{\hat{\rho}}}_{0}\equiv\boldsymbol{\mathrm{\hat{\rho}}}_{s}(0)$,
and operate $K$ times with $\mathcal{L}$ to get the set 
$\{\boldsymbol{\mathrm{\hat{\rho}}}_{0},\allowbreak
\mathcal{L}\boldsymbol{\mathrm{\hat{\rho}}}_{0},\allowbreak
\mathcal{L}^{2}\boldsymbol{\mathrm{\hat{\rho}}}_{0},\allowbreak
\ldots,\allowbreak
\mathcal{L}^{K}\boldsymbol{\mathrm{\hat{\rho}}}_{0}\}$
which is the starting point for orthogonalization and $(K+1)\times(K+1)$-dimension
matrix representation of $\mathcal{L}$. We note that the operation
of $\mathcal{L}$ involves $n\times n$ matrix-matrix multiplications,
which scales as $n^{3}$. Therefore, it is more efficient to use the
operation of $\mathcal{L}$ for the procedure than to use the vec-ing
matrix $L$ (Eq. (\ref{eq:L_vec_ing})) described in Sec. \ref{sec:vec_theory}
above.

The actual procedure follows, adapted to the notation of a super-operator
and density matrices:
\begin{enumerate}
\item Begin with the normalized density matrix $\boldsymbol{\mathrm{\hat{\rho}}}_{0}$.
\item for $j=0$ to $K$

\begin{enumerate}
\item Compute a non-orthonormalized new density matrix by setting: $\boldsymbol{\mathrm{\hat{\rho}}}_{j+1}:=\mathcal{L}\boldsymbol{\mathrm{\hat{\rho}}}_{j}$
\item for $i=0$ to $j$

\begin{enumerate}
\item Set: $L_{i,j}:=\left(\boldsymbol{\mathrm{\hat{\rho}}}_{i}^{\dagger},\boldsymbol{\mathrm{\hat{\rho}}}_{j+1}\right)=\text{Tr}\left\{ \boldsymbol{\mathrm{\hat{\rho}}}_{i}^{\dagger}\boldsymbol{\mathrm{\hat{\rho}}}_{j+1}\right\} $
\item Subtract the projection on $\boldsymbol{\mathrm{\hat{\rho}}}_{i}$:
$\boldsymbol{\mathrm{\hat{\rho}}}_{j+1}:=\boldsymbol{\mathrm{\hat{\rho}}}_{j+1}-L_{i,j}\boldsymbol{\mathrm{\hat{\rho}}}_{i}$
\end{enumerate}
\item end for
\item Set: $L_{j+1,j}:=\left\Vert \boldsymbol{\mathrm{\hat{\rho}}}_{j+1}\right\Vert \equiv\sqrt{\text{Tr}\left\{ \boldsymbol{\mathrm{\hat{\rho}}}_{j+1}^{\dagger}\boldsymbol{\mathrm{\hat{\rho}}}_{j+1}\right\} }$
\item Normalize $\boldsymbol{\mathrm{\hat{\rho}}}_{j+1}$ by setting $\boldsymbol{\mathrm{\hat{\rho}}}_{j+1}:=\frac{\boldsymbol{\mathrm{\hat{\rho}}}_{j+1}}{L_{j+1,j}}$
\end{enumerate}
\item end for
\end{enumerate}
The procedure yields
\[
L_{i,j} = 
\text{Tr}\left\{ \boldsymbol{\mathrm{\hat{\rho}}}_{i}^{\dagger}\mathcal{L}\boldsymbol{\mathrm{\hat{\rho}}}_{j}\right\}  \qquad i\le j+1
\]
For \text{$i>j+1$}, the expression in the RHS vanishes. 
Thus, we can define a $(K+1)\times(K+1)$ matrix which its general element is given by a matrix element of $\mathcal{L}$ in the Liouville space:
\[
L_{i,j} = 
\text{Tr}\left\{ \boldsymbol{\mathrm{\hat{\rho}}}_{i}^{\dagger}\mathcal{L}\boldsymbol{\mathrm{\hat{\rho}}}_{j}\right\}  
\]
(Note that the procedure also yields $\boldsymbol{\mathrm{\hat{\rho}}}_{K+1}$
and $L_{K+1,K}$ which are not necessary for our purposes). $L$ represents
the operation of the super-operator $\mathcal{L}$ on the subspace
that is spanned by the density matrices $\{\boldsymbol{\mathrm{\hat{\rho}}}_{0},\boldsymbol{\mathrm{\hat{\rho}}}_{1},\boldsymbol{\mathrm{\hat{\rho}}}_{2},\ldots,\boldsymbol{\mathrm{\hat{\rho}}}_{K}\}$.
The matrix $L$ is referred to as the \emph{Hessenberg matrix of} $\mathcal{L}$.
The density matrix has to be approximated by
its projection on the subspace: $\boldsymbol{\mathrm{\hat{\rho}}}\approx r_{0}\boldsymbol{\mathrm{\hat{\rho}}}_{0}+r_{1}\boldsymbol{\mathrm{\hat{\rho}}}_{1}+r_{2}\boldsymbol{\mathrm{\hat{\rho}}}_{2}+\ldots+r_{K}\boldsymbol{\mathrm{\hat{\rho}}}_{K}$.
The vector 
\begin{equation}
\vec{r}\equiv(r_{0},r_{1},r_{2},\ldots,r_{K})^{T}\label{eq:vector_arnoldi}
\end{equation}
 is the representation of the density matrix in this subspace. The
dynamics of the vector $\vec{r}$ is generated by the matrix $L$
that was constructed in step (2) of the above procedure:
\[
\frac{d}{dt}\vec{r}=L\vec{r}.
\]
 Exponentiation and eigenvalue calculations of the matrix $L$ can
be done by common numerical techniques \cite{arnoldi1951principle,saad1992analysis_krylov}.

The Arnoldi algorithm usually becomes problematic when a large dimension approximation is required, i.e.\ when $K$ is large. In such a case, a \emph{restarted Arnoldi algorithm} should be used instead (see, for example, \cite{restartedArnoldi}). This topic is beyond the scope of this paper.

\subsection{The Heisenberg representation}
\par
Not always the full state of the system will be of concern. 
In most cases we will be interested only in the expectation values
of some measured quantities. 
This fact can reduce significantly the dimensions of the problem. 
For example, in the standard thermalizing master equation the population and the coherences are decoupled, and the population of a certain level is given by solving a single differential equation \cite{breuer}. 
The full state of the system can be reconstructed by calculating all the expectation values of the Lie algebra of the system. 
Generally, a full reconstruction of the state will scale as the Vec-ing of the density matrix introduced in Sec. \ref{sec:vec_theory}. 
Nevertheless, in many cases we can use symmetries to reduce the dimensions of the problem. 
For example, if the initial state of harmonic oscillator is a Gaussian state, then it will stay Gaussian along the dynamics and only the first two moments are necessary to retrieve the full state
\cite{rezek2006}. 
Another example is coupled two qubits in which the full dimension of the system is 16, 
but only 3 operators are sufficient to define the energy and coherence of the system 
\cite{tova2002discrete}. 
\par                
To describe the dynamics of the expectation values, it is common
to use the master equation in the Heisenberg representation. The operator
$\boldsymbol{\mathrm{\hat{X}}}$ belonging to dual Hilbert space of
the system follows the dynamics \cite{alicki2001quantum,breuer}:
\begin{equation}
\boldsymbol{\mathrm{\hat{X}}}(t)=e^{\mathcal{L^{\dagger}}t}\boldsymbol{\mathrm{\hat{X}}}(0),
\end{equation}
which in its differential form is written explicitly as 
\begin{equation}
\frac{d}{dt}\boldsymbol{\mathrm{\hat{X}}} 
=
\mathcal{L}^\dagger \boldsymbol{\mathrm{\hat{X}}} 
\equiv 
\frac{i}{\hbar}\left[\boldsymbol{\mathrm{\hat{H}}}, \boldsymbol{\mathrm{\hat{X}}}\right]
+ 
\sum_{i}\gamma_{i}\left(
\boldsymbol{\mathrm{\hat{A}}}_{i}^{\dagger} \boldsymbol{\mathrm{\hat{X}}} \boldsymbol{\mathrm{\hat{A}}}_{i}
-
\frac{1}{2}\left\lbrace 
\boldsymbol{\mathrm{\hat{A}}}_{i}^{\dagger} \boldsymbol{\mathrm{\hat{A}}}_{i}, 
\boldsymbol{\mathrm{\hat{X}}}\right\rbrace \right).
\label{eq:LGKS_heis}
\end{equation}

If there is a a set of operators $\{\boldsymbol{\mathrm{\hat{X}}}_{k}\}_{k=1}^{M}$,
$M<n^2$, that forms a closed set under the operation of ${\cal L^{\dagger}}$,
meaning 
\begin{equation}
\mathcal{L}^{\dagger}\boldsymbol{\mathrm{\hat{X}}}_{k}
=
\sum_{j=1}^{M} l_{kj} \boldsymbol{\mathrm{\hat{X}}}_{j}
\label{eq:heisenberg_set}
\end{equation}
then we can write a closed linear system of coupled differential equations.
The expectation values $x_{k}\equiv\left\langle \boldsymbol{\mathrm{\hat{X}}}_{k}\right\rangle $
will have the corresponding set of coupled differential equations.
The analytical form of their dynamics will follow the form of Eq.
(\ref{eq:sum_exp}). We define the vector of expectation values $\vec{R}\equiv(x_{1},x_{2},\ldots)^{T}$.
This system can be represented in a matrix-vector notation, 

\[
\frac{d}{dt}\vec{R}=L^{\dagger}\vec{R},
\]
where the matrix $L^\dagger$ is defined by the equation set Eq. (\ref{eq:heisenberg_set}), $\left( L^\dagger \right)_{kj}=l_{kj}$.
The dimension of this matrix is $M^{2}$. 
Note that eigenvalues of the matrix $L^\dagger$ are complex conjugates of a subset of the eigenvalues of the
super-operator $\mathcal{L}$ of Eq. (\ref{eq:LGKS_schrod}).

\section{Example: The two-level system master equation}

\subsection{The model}

As an example, we consider a driven two-level system (TLS) with spontaneous emission \cite{agarwal1970master,cohen1998optical}.

We use the following definitions:

\[
\begin{array}{cccc}
\boldsymbol{\mathrm{\hat{S}}}_{x}=\frac{1}{2}\left(\begin{array}{cc}
0 & 1\\
1 & 0
\end{array}\right), & \boldsymbol{\mathrm{\hat{S}}}_{y}=\frac{1}{2}\left(\begin{array}{cc}
0 & -i\\
i & 0
\end{array}\right), & \boldsymbol{\mathrm{\hat{S}}}_{z}=\frac{1}{2}\left(\begin{array}{cc}
1 & 0\\
0 & -1
\end{array}\right), & \boldsymbol{\mathrm{\hat{I}}}=\left(\begin{array}{cc}
1 & 0\\
0 & 1
\end{array}\right)\end{array}.
\]
In addition:
\[
\begin{array}{cc}
\boldsymbol{\mathrm{\hat{S}}}_{+}\equiv\boldsymbol{\mathrm{\hat{S}}}_{x}+i\boldsymbol{\mathrm{\hat{S}}}_{y}=\left(\begin{array}{cc}
0 & 1\\
0 & 0
\end{array}\right), & \boldsymbol{\mathrm{\hat{S}}}_{-}\equiv\boldsymbol{\mathrm{\hat{S}}}_{x}-i\boldsymbol{\mathrm{\hat{S}}}_{y}=\left(\begin{array}{cc}
0 & 0\\
1 & 0
\end{array}\right)\end{array}.
\]
The commutation relations are:
\[
\left[\boldsymbol{\mathrm{\hat{S}}}_{i},\boldsymbol{\mathrm{\hat{S}}}_{j}\right]=i\epsilon_{ijk}\boldsymbol{\mathrm{\hat{S}}}_{k},
\]
where $\epsilon_{ijk}$ is the Levi-Civita symbol, defined as:

\[
\epsilon_{ijk}
= \left\lbrace
\begin{array}{cl}
+1 & ijk\text{ is cyclic permutation of }xyz\\
-1 & ijk\text{ is anti-cyclic permutation of }xyz\\
0 & \text{\ensuremath{i=j} or \ensuremath{j=k} or \ensuremath{k=i}}.
\end{array}
\right.
\]

The system Hamiltonian is 
\[
\boldsymbol{\mathrm{\hat{H}}}_{S}=\omega_{s}\boldsymbol{\mathrm{\hat{S}}}_{z},
\]
with $\omega_{s}$ as the system transition frequency. The system
is driven by the external field $f(t)=\varepsilon e^{-i\omega_{L}t}$,
with the carrier frequency $\omega_{L}$ and amplitude $\varepsilon$.
The coupling to the driving field is expressed by the matrix
\[
\boldsymbol{\mathrm{\hat{V}}}=\left(\begin{array}{cc}
0 & f(t)\\
f^{*}(t) & 0
\end{array}\right).
\]
In order to work within a time-independent Hamiltonian, we move to
an interaction picture according to $\omega_{L}\boldsymbol{\mathrm{\hat{S}}}_{z}$,
obtaining 
\[
\boldsymbol{\mathrm{\hat{H}}}=\Delta\boldsymbol{\mathrm{\hat{S}}}_{z}+\varepsilon\boldsymbol{\mathrm{\hat{S}}}_{x},
\]
where $\Delta\equiv\omega_{s}-\omega_{L}$ is the detuning between
the system and the field frequencies.

The spontaneous emission is expressed by a dissipative term, and the
L-GKS master equation takes the form \cite{agarwal1970master,cohen1998optical}:

\begin{equation}
\frac{d}{dt}\boldsymbol{\mathrm{\hat{\rho}}}_{S}=-i\left[\boldsymbol{\mathrm{\hat{H}}},\boldsymbol{\mathrm{\hat{\rho}}}_{S}\right]+\gamma\left(\boldsymbol{\mathrm{\hat{S}}}_{-}\boldsymbol{\mathrm{\hat{\rho}}}_{S}\boldsymbol{\mathrm{\hat{S}}}_{+}-\frac{1}{2}\left\{ \boldsymbol{\mathrm{\hat{S}}}_{+}\boldsymbol{\mathrm{\hat{S}}}_{-},\boldsymbol{\mathrm{\hat{\rho}}}_{S}\right\} \right),\label{eq:tls_lgks}
\end{equation}
where $\gamma$ is the spontaneous emission rate.

\subsection{Matrix-vector representation of the TLS dynamics}

In the following we will implement the different approaches discussed above, 
for the relaxed driven TLS.

\subsubsection{Vec-ing the density matrix}

We use the procedure presented in Sec. (\ref{sec:vec_theory}) for the L-GKS
of the two-level system with relaxation, Eq. (\ref{eq:tls_lgks}). 

The basis for the representation is the trivial set of matrices: 
\[
\left\{ \begin{array}{cccc}
\left(\begin{array}{cc}
0 & 0\\
0 & 1
\end{array}\right), & \left(\begin{array}{cc}
0 & 0\\
1 & 0
\end{array}\right), & \left(\begin{array}{cc}
0 & 1\\
0 & 0
\end{array}\right), & \left(\begin{array}{cc}
0 & 0\\
0 & 1
\end{array}\right)\end{array}\right\} .
\]

To represent the commutator as a $n^{2}\times n^{2}=4\times4$ matrix,
we use the procedure to get:
\begin{eqnarray*}
-i\Delta\left(I\otimes\boldsymbol{\mathrm{\hat{S}}}_{z}-\boldsymbol{\mathrm{\hat{S}}}_{z}^{T}\otimes I\right) & = & -i\left(\begin{array}{cccc}
0 & 0 & 0 & 0\\
0 & \Delta & 0 & 0\\
0 & 0 & -\Delta & 0\\
0 & 0 & 0 & 0
\end{array}\right)\\
-i\varepsilon\left(I\otimes\boldsymbol{\mathrm{\hat{S}}}_{x}-\boldsymbol{\mathrm{\hat{S}}}_{x}^{T}\otimes I\right) & = & -\frac{i}{2}\left(\begin{array}{cccc}
0 & \varepsilon & -\varepsilon & 0\\
\varepsilon & 0 & 0 & -\varepsilon\\
-\varepsilon & 0 & 0 & \varepsilon\\
0 & -\varepsilon & \varepsilon & 0
\end{array}\right).
\end{eqnarray*}
For the dissipator we have:
\[
\gamma\left(\boldsymbol{\mathrm{\hat{S}}}_{+}\right)^{T}\otimes\boldsymbol{\mathrm{\hat{S}}}_{-}-\frac{\gamma}{2}\left(I\otimes\left(\boldsymbol{\mathrm{\hat{S}}}_{+}\boldsymbol{\mathrm{\hat{S}}}_{-}\right)+\left(\boldsymbol{\mathrm{\hat{S}}}_{+}\boldsymbol{\mathrm{\hat{S}}}_{-}\right)^{T}\otimes I\right)=\gamma\left(\begin{array}{cccc}
-1 & 0 & 0 & 0\\
0 & -\frac{1}{2} & 0 & 0\\
0 & 0 & -\frac{1}{2} & 0\\
1 & 0 & 0 & 0
\end{array}\right).
\]
We combine all parts to get:
\begin{equation}
L=\left(\begin{array}{cccc}
-\gamma & -i\frac{\varepsilon}{2} & i\frac{\varepsilon}{2} & 0\\
-i\frac{\varepsilon}{2} & -\frac{1}{2}\gamma-i\Delta & 0 & i\frac{\varepsilon}{2}\\
i\frac{\varepsilon}{2} & 0 & -\frac{1}{2}\gamma+i\Delta & -i\frac{\varepsilon}{2}\\
\gamma & i\frac{\varepsilon}{2} & -i\frac{\varepsilon}{2} & 0
\end{array}\right).
\end{equation}

\subsubsection{Arnoldi method}

The Arnoldi method that was presented in Sec. \ref{sec:arnoldi_theory},
and we use it here to find a representation of the two-level system
super-operator. Note that generally the Arnoldi method is used for
the \emph{approximation of large matrices}. Here we have a small-size
problem ($n^{2}=4$), and we create an \emph{exact} matrix, which
represents the two-level system super-operator on a basis that spans
the entire space. 

We start with an initial density matrix $\rho_{0}$ (chosen arbitrary):
\[
\rho_{0}=\left(\begin{array}{cc}
0 & 0\\
0 & 1
\end{array}\right).
\]
Then we follow the Arnoldi iteration procedure to get the basis $\{\boldsymbol{\mathrm{\hat{\rho}}}_{0},\boldsymbol{\mathrm{\hat{\rho}}}_{1},\boldsymbol{\mathrm{\hat{\rho}}}_{2},\boldsymbol{\mathrm{\hat{\rho}}}_{3}\}$:
\[
\left\{ \begin{array}{cccc}
\left(\begin{array}{cc}
0 & 0\\
0 & 1
\end{array}\right), & \frac{1}{\sqrt{2}}\left(\begin{array}{cc}
0 & -i\\
i & 0
\end{array}\right), & \frac{1}{\Omega}\left(\begin{array}{cc}
\varepsilon & \Delta\\
\Delta & 0
\end{array}\right), & \frac{1}{\sqrt{2}\Omega}\left(\begin{array}{cc}
-2\Delta & \varepsilon\\
\varepsilon & 0
\end{array}\right)\end{array}\right\} ,
\]
with the definition: $\Omega=\sqrt{2\Delta^{2}+\varepsilon^{2}}$. This
implies that the initial vector in this basis is $\vec{r}(0)=(1,0,0,0)^{T}$.

The representation of the super-operator $\mathcal{L}$ in this basis
is the matrix obtained by the procedure:
\begin{equation}
L=\left(\begin{array}{cccc}
0 & -\frac{\varepsilon}{\sqrt{2}} & \frac{\varepsilon\gamma}{\Omega} & -\frac{\sqrt{2}\gamma\Delta}{\Omega}\\
\frac{\varepsilon}{\sqrt{2}} & -\frac{\gamma}{2} & -\frac{\Omega}{\sqrt{2}} & 0\\
0 & \frac{\Omega}{\sqrt{2}} & -\frac{\gamma\left(\Delta^{2}+\varepsilon^{2}\right)}{\Omega^{2}} & \frac{\gamma\Delta\varepsilon}{\sqrt{2}\Omega^{2}}\\
0 & 0 & \frac{\Delta\varepsilon\gamma}{\sqrt{2}\Omega^{2}} & \frac{\gamma\varepsilon^{2}}{2\Omega^{2}}-\gamma
\end{array}\right).
\end{equation}

\subsubsection{The Heisenberg representation}

For the system described above, the L-GKS equation in the Heisenberg picture is:

\[
\frac{d}{dt}\boldsymbol{\mathrm{\hat{X}}}
=
i\left[\boldsymbol{\mathrm{\hat{H}}}, \boldsymbol{\mathrm{\hat{X}}}\right]
+
\gamma\left(
\boldsymbol{\mathrm{\hat{S}}}_{+} \boldsymbol{\mathrm{\hat{X}}} \boldsymbol{\mathrm{\hat{S}}}_{-}
-
\frac{1}{2}\left\{ 
\boldsymbol{\mathrm{\hat{S}}}_{+} \boldsymbol{\mathrm{\hat{S}}}_{-}, \boldsymbol{\mathrm{\hat{X}}}
\right\} 
\right).
\]
As a set of basis operators we choose the set: 
\[
\left\{ \begin{array}{cccc}
\boldsymbol{\mathrm{\hat{S}}}_{x}, & \boldsymbol{\mathrm{\hat{S}}}_{y}, & \boldsymbol{\mathrm{\hat{S}}}_{z}, & \boldsymbol{\mathrm{\hat{I}}}\end{array}\right\} .
\]
The L-GKS equation for the operators of the basis gives:
\[
\begin{array}{ccccc}
\frac{d}{dt}\boldsymbol{\mathrm{\hat{S}}}_{x} & = & i\left[\boldsymbol{\mathrm{\hat{H}}},\boldsymbol{\mathrm{\hat{S}}}_{x}\right] & + & \gamma\left(\boldsymbol{\mathrm{\hat{S}}}_{+}\boldsymbol{\mathrm{\hat{S}}}_{x}\boldsymbol{\mathrm{\hat{S}}}_{-}-\frac{1}{2}\left\{ \boldsymbol{\mathrm{\hat{S}}}_{+}\boldsymbol{\mathrm{\hat{S}}}_{-},\boldsymbol{\mathrm{\hat{S}}}_{x}\right\} \right)\\
 & = & -\Delta\boldsymbol{\mathrm{\hat{S}}}_{y} & - & \frac{1}{2}\gamma\boldsymbol{\mathrm{\hat{S}}}_{x},
\end{array}
\]
\[
\begin{array}{ccccc}
\frac{d}{dt}\boldsymbol{\mathrm{\hat{S}}}_{y} & = & i\left[\boldsymbol{\mathrm{\hat{H}}},\boldsymbol{\mathrm{\hat{S}}}_{y}\right] & + & \gamma\left(\boldsymbol{\mathrm{\hat{S}}}_{+}\boldsymbol{\mathrm{\hat{S}}}_{y}\boldsymbol{\mathrm{\hat{S}}}_{-}-\frac{1}{2}\left\{ \boldsymbol{\mathrm{\hat{S}}}_{+}\boldsymbol{\mathrm{\hat{S}}}_{-},\boldsymbol{\mathrm{\hat{S}}}_{y}\right\} \right)\\
 & = & \Delta\boldsymbol{\mathrm{\hat{S}}}_{y}-\varepsilon\boldsymbol{\mathrm{\hat{S}}}_{z} & - & \frac{1}{2}\gamma\boldsymbol{\mathrm{\hat{S}}}_{y},
\end{array}
\]
\[
\begin{array}{ccccc}
\frac{d}{dt}\boldsymbol{\mathrm{\hat{S}}}_{z} & = & i\left[\boldsymbol{\mathrm{\hat{H}}},\boldsymbol{\mathrm{\hat{S}}}_{z}\right] & + & \gamma\left(\boldsymbol{\mathrm{\hat{S}}}_{+}\boldsymbol{\mathrm{\hat{S}}}_{z}\boldsymbol{\mathrm{\hat{S}}}_{-}-\frac{1}{2}\left\{ \boldsymbol{\mathrm{\hat{S}}}_{+}\boldsymbol{\mathrm{\hat{S}}}_{-},\boldsymbol{\mathrm{\hat{S}}}_{z}\right\} \right)\\
 & = & \varepsilon\boldsymbol{\mathrm{\hat{S}}}_{y} & - & \gamma\boldsymbol{\mathrm{\hat{S}}}_{z}-\frac{1}{2}\gamma\boldsymbol{\mathrm{\hat{I}}},
\end{array}
\]
and, of course: 
\[
\frac{d}{dt}\boldsymbol{\mathrm{\hat{I}}}=0.
\]

The expectation values of these operators, will follow the same dynamics.
We denote the vector of this expectation values as $\vec{R}\equiv\{s_{x},s_{y},s_{z},I\}$.
The dynamics of this vector is given by 
\[
\frac{d}{dt}\vec{R}=L^{\dagger}\vec{R},
\]
with the matrix:
\begin{equation}
L^{\dagger}=\left(\begin{array}{cccc}
-\frac{1}{2}\gamma & -\Delta & 0 & 0\\
\Delta & -\frac{1}{2}\gamma & -\varepsilon & 0\\
0 & \varepsilon & -\gamma & -\frac{1}{2}\gamma\\
0 & 0 & 0 & 0
\end{array}\right).
\end{equation}

\subsection{Comparison}
We demonstrated the three approaches for representing the L-GKS dynamics with a matrix-vector notation. 
For the relaxed two-level system we obtained three different matrices. 
However, they are all equivalent, representing the dynamics by different bases or in different spaces. 
A simple verification is calculating the eigenvalues for different values of the parameters. 
As a neat example, for the values of 
$\Delta=\sqrt{\frac{1}{108}}\gamma$,
$\varepsilon=\sqrt{\frac{8}{108}}\gamma$, 
we know from a previous study \cite{morag2014EPbloch} 
that we get a third order non-hermitian degeneracy of the eigenvalue $\lambda=-\frac{2}{3}\gamma$.
Substituting these values to the three matrices yield the same non-hermitian degeneracy.

\section{Discussion}

There is a heavy conceptual and computational price for the reduced description of open quantum systems in Liouville space. 
To pave the way to overcome this difficulty, it is desirable to represent the dynamics in the more familiar matrix vector notation.

Significant simplification can be identified in the Heisenberg representation, 
when a set of operators which is closed to the equation of motion is found. 
For the Hamiltonian part, a closed set is obtained when the operators form a closed compact Lie algebra and the Hamiltonian is a linear combination of these operators \cite{alhassid1978}.
Additional requirements are needed for the set to be also closed to the dissipative part \cite[Appendix A]{rezek2006}.

When a closed set of operators cannot be found one has to resort to approximate methods. 
The idea is to construct a representative subset of operators. 
This set is generated from the initial state with successive applications of the dynamical generator ${\cal L}$. 
The initial idea can be traced to Lanczos \cite{lanczos1950iteration} who applied it to obtain iterative solutions to eigenvalue problems
of a hermitian operator $\mathbf A$. 
Since the eigenvalues of ${\cal L}$ are complex, 
the iterative approach is modified to the Arnoldi method \cite{arnoldi1951principle}.
The Arnoldi approach is effective in the reduction of a large scale problem into a relatively small approximation space.
Therefore, it should be considered as a standard approach for the treatment of large-scale L-GKS problems.

\subsection*{Bibliography}
\bibliographystyle{unsrt}
\bibliography{three_methods}

\end{document}